# Rotational Dynamics of Organic Cations in $CH_3NH_3PbI_3$ Perovskite

Tianran Chen[1,†], Benjamin J. Foley[2,†], Bahar Ipek[3,4], Madhusudan Tyagi[3], John R. D. Copley[3], Craig M. Brown[3,4], Joshua J. Choi[2]*, Seung-Hun Lee[1]*

[1] Department of Physics, University of Virginia, Charlottesville, Virginia 22904, USA.

[2] Department of Chemical Engineering, University of Virginia, Charlottesville, Virginia 22904, USA.

[3] NIST Center for Neutron Research, National Institute of Standards and Technology, Gaithersburg, Maryland 20899, USA.

[4] Department of Chemical and Biomolecular Engineering, University of Delaware, Newark, Delaware 19716, USA.

†These authors contributed equally to this work.

*Corresponding authors. E-mail: shlee@virginia.edu, jjc6z@virginia.edu

## Abstract

Methylammonium lead iodide ($CH_3NH_3PbI_3$) based solar cells have shown impressive power conversion efficiencies of above 20%. However, the microscopic mechanism of the high photovoltaic performance is yet to be fully understood. Particularly, the dynamics of $CH_3NH_3^+$ cations and their impact on relevant processes such as charge recombination and exciton dissociation are still poorly understood. Here, using elastic and quasi-elastic neutron scattering techniques and group theoretical analysis, we studied rotational modes of the $CH_3NH_3^+$ cation in $CH_3NH_3PbI_3$. Our results show that, in the cubic (T > 327K) and tetragonal (165K < T < 327K) phases, the $CH_3NH_3^+$ ions exhibit four-fold rotational symmetry of the C-N axis ($C_4$) along with three-fold rotation around the C-N axis ($C_3$), while in orthorhombic phase (T < 165K) only $C_3$ rotation is present. Around room temperature, the characteristic relaxation times for the $C_4$ rotation is found to be $\tau_{C_4} \approx 5$ ps while for the $C_3$ rotation $\tau_{C_3} \approx 1$ ps. The $T$-dependent rotational relaxation times were fitted with Arrhenius equations to obtain activation energies. Our data show a close correlation between the $C_4$ rotational mode and the temperature dependent dielectric permittivity. Our findings on the rotational dynamics of $CH_3NH_3^+$ and the associated dipole have important implications on understanding the low exciton binding energy and slow charge recombination rate in $CH_3NH_3PbI_3$ which are directly relevant for the high solar cell performance.

**Introduction**

Organic-inorganic hybrid perovskites are transforming the solar cell research field, having rapidly reached photovoltaic power conversion efficiency of above 20 %[1, 2] after only five years since the first report.[3] High photovoltaic efficiency requires three ingredients; bandgap well-matched with the solar spectrum, large light absorption coefficient and long charge carrier diffusion length. The most extensively studied hybrid perovskite, $CH_3NH_3PbI_3$, has been shown to possess these three features; bandgap of 1.6 eV,[4, 5] large extinction coefficient of greater than $10^4$ $cm^{-1}$,[6] and a long electron-hole diffusion length in the range of hundreds of nanometers for polycrystalline thin films to greater than 175 microns for single crystals.[7-11] In addition to these favorable characteristics, $CH_3NH_3PbI_3$ exhibits several peculiar optoelectronic properties that are relevant to photovoltaic performance - giant increase (3 orders of magnitude) in dielectric constant upon light illumination,[12] extremely slow photoconductivity response[13] and possible existence of ferroelectric domains.[14] The microscopic mechanism of the high photovoltaic performance and these properties is yet to be fully understood.

Among proposed microscopic pictures, one scenario is based on presence of nanoscale ferroelectric domains due to alignment of organic $CH_3NH_3^+$ cations with electric dipole[15-19] and rearrangement of the inorganic scaffold that is intimately linked with the $CH_3NH_3^+$ ion orientation.[13, 19-21] The ferroelectric domains provide internal electric field and separate photogenerated electrons and holes, thus reducing the probability of their recombination. This has been invoked to explain the slow charge recombination[22, 23] and long charge diffusion lengths[10, 11] However, ferroelectricity in $CH_3NH_3PbI_3$ is still being actively debated - while some studies indicate presence of ferroelectricity,[14, 24-26] other studies have found that $CH_3NH_3PbI_3$ does not show permanent electric polarization at room temperature.[27, 28] Another microscopic mechanism currently being investigated is the ionic conduction where charged species readily move in $CH_3NH_3PbI_3$ to create space charge regions.[27, 29-32] This mechanism is likely to be the main cause of the current-voltage hysteresis observed in some solar cells,[33] with a strong evidence recently obtained from switchable electric polling in planar device that is independent of $CH_3NH_3PbI_3$ layer thickness.[32] The ion conduction mechanism has also been invoked to explain some of the peculiar dielectric behaviors mentioned above. These different microscopic pictures may have varying significance depending on the properties under consideration and sample preparation methods.

Whether the ferroelectricity in $CH_3NH_3PbI_3$ at room temperature exists or not, the orientation and dynamics of $CH_3NH_3^+$ ions are thought to critically impact various properties relevant for solar cell performance such as light-induced giant dielectric constant increase,[12] picosecond dielectric relaxation,[34] slow charge recombination,[23, 35] electronic structure[18, 35] and low (few meV) exciton binding energy.[36-38] Currently, there is no consensus on the microscopic mechanism for these behaviors due to lack of full understanding of the atomic structure and dynamics of $CH_3NH_3PbI_3$. Recently, Leguy *et al.*[39] have reported quasi-elastic neutron scattering studies with limited data range and analysis which led to results that are inconsistent with previous experimental results obtained by millimeter-wave spectroscopy,[34] nuclear magnetic resonance,[40, 41] as well as theoretical prediction with *ab initio* molecular dynamics simulations.[18, 35, 42] This calls for accurate and in-depth studies on the dynamics of $CH_3NH_3^+$ ions to advance understanding of the rich behavior of $CH_3NH_3PbI_3$.

We have performed elastic and quasi-elastic neutron scattering to probe rotational motions in $CH_3NH_3PbI_3$ covering the full range of relevant regions in energy ($\hbar\omega$) and momentum ($Q$) phase space over a wide range of temperature, $370 \text{ K} \geq T \geq 70 \text{ K}$, covering the three different structural phases of $CH_3NH_3PbI_3$. Our comprehensive data enabled us to employ a group theoretical method based on the crystal symmetry to understand the nature of the rotations of the $CH_3NH_3^+$ cation, including their symmetries, the relaxation times as a function of temperature, and the activation energies. Our results show that, in the cubic ($T > 327 \text{ K}$) and tetragonal ($165 \text{ K} < T < 327 \text{ K}$) phases, the $CH_3NH_3^+$ ions exhibit four-fold rotational symmetry of the C-N axis ($C_4$) along with three-fold rotation around the C-N axis ($C_3$), while in orthorhombic phase ($T < 165 \text{ K}$) only $C_3$ rotation is present. We show that the onset of $C_4$ rotation at 160K, upon heating, is well correlated with bulk properties such as sudden jump in dielectric permittivity.[43] Around room temperature, the characteristic relaxation times for the $C_4$ rotation is found to be $\tau_{C_4} \approx 5$ ps while for the $C_3$ rotation $\tau_{C_3} \approx 1$ ps. The $T$-dependent rotational relaxation times were fitted with Arrhenius equations to obtain activation energies. Intriguing shifts in activations energies as a function of crystal symmetry were found.

**Experimental**

**Materials.** The following chemicals were used as received: 33% methylamine in ethanol, hydroiodic acid (57% by weight in water), ethanol, diethyl ether, lead iodide ($PbI_2$) (99.999%) from Sigma Aldrich and cesium iodide (CsI) (99.999%) from Alfa Aesar. Methylammonium iodide was synthesized by following a method in the literature.[44] $CH_3NH_3PbI_3$ and $CsPbI_3$ were synthesized via solution crystallization by mixing methylammonium iodide or CsI with $PbI_2$ in aqueous hydrogen iodide solution and slowly evaporating the liquids via heating in ambient air. 13.5g of $CH_3NH_3PbI_3$ and 16.4g of $CsPbI_3$ powder samples were obtained for neutron scattering measurement.

**Neutron Scattering Measurements**. Crystal structure and structural parameters at 4 K were determined by neutron powder diffraction (NPD), using the BT1 diffractometer at the NIST Center for Neutron Research (NCNR) located in Gaithersburg, Maryland, USA. The wavelength was selected using a Cu(311) monochromator with an in-pile 60' collimator ( $\lambda = 1.5398\,\text{Å}$ ). The scattered neutrons were collected by 32 3-He detectors over the 2θ range of 1.3°– 166.3° with 0.05° step size. $CH_3NH_3PbI_3$ (8 g) sample was placed into a cylindrical vanadium can in a dry helium box. The vanadium can was sealed with an indium o-ring. The residual helium inside the vanadium can was removed using a turbo molecular pump prior to diffraction experiments, and the sample can was mounted in a closed cycle helium refrigerator.

Time-of-flight neutron scattering measurements on the polycrystalline sample of $CH_3NH_3PbI_3$ on the Disk-Chopper-Spectrometer (DCS)[45] and the High Flux Backscattering Spectrometer (HFBS)[46] were performed at the NCNR. At DCS, energy of incident neutrons was fixed to $E_i = 3.55$ meV (wavelength $\lambda = 4.8\,\text{Å}$) with an energy resolution of 0.1 meV at the elastic channel. At HFBS, energy of scattered neutrons was fixed to $E_f = 2.08$ meV ($l = 6.271$ Å) with an energy resolution of 1 µeV. In both experiments, the sample was mounted in a closed cycle helium refrigerator.

## Results

### Quasi-elastic Scattering from $CH_3NH_3PbI_3$

Fig. 1 shows quasi-elastic scattering from $CH_3NH_3PbI_3$ measured at four different temperatures with DCS. At 370 K in the cubic phase, there is a broad peak centered at $Q \sim 1.5$ Å$^{-1}$ and $\hbar\omega = 0$ meV. Upon cooling into tetragonal phase, as shown at 260 K and 180 K, the peak becomes narrow in energy resulting in the strong signal at low energies. Upon further cooling to 130 K in orthorhombic phase, the strong low energy signal moves to much lower energies and most of it falls into the energy window of the instrumental energy resolution of 0.1 meV. The color-coded surfaces are the results of two-dimensional global fits of the data with models of rotational motions of the $CH_3NH_3^+$ molecule in the respective local environments of the crystal structures, which will be described in detail later.

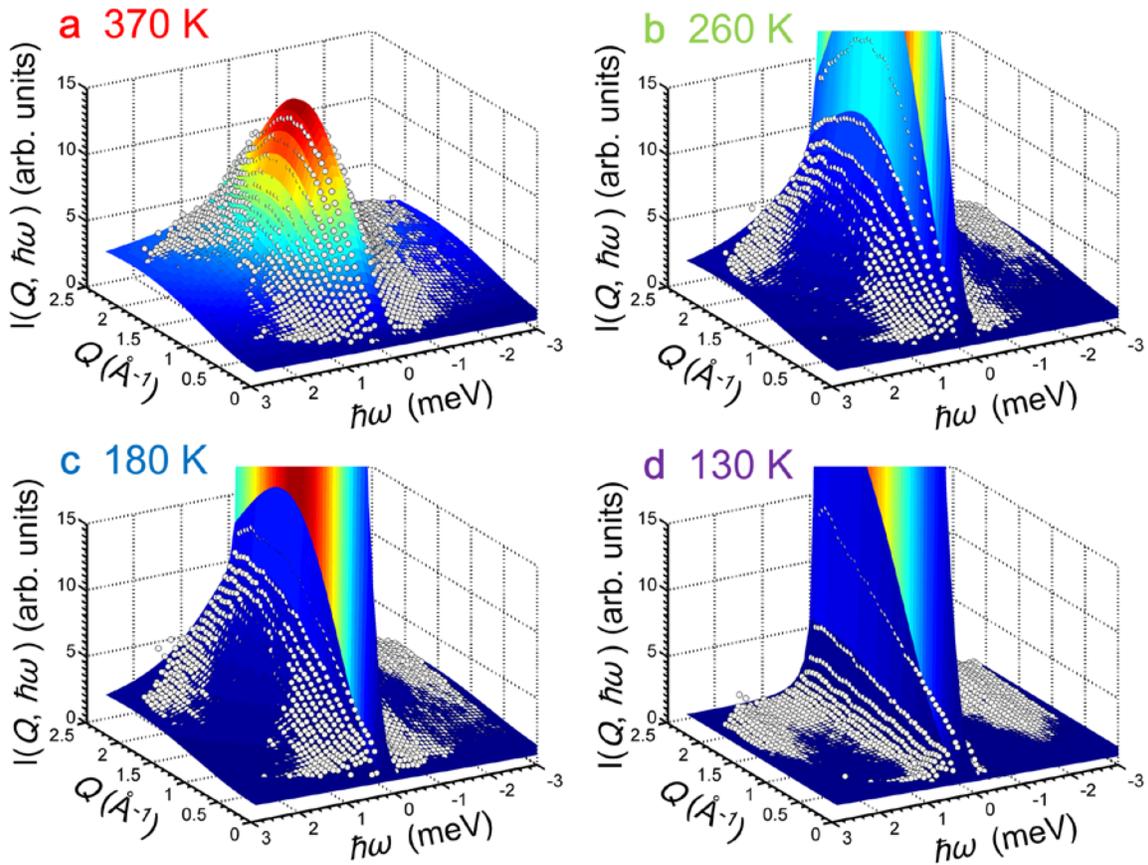

**Figure 1. Quasi-elastic scattering from CH$_3$NH$_3$PbI$_3$.** Neutron scattering intensity is shown as a function of momentum ($Q$) and energy ($\hbar\omega$) transfers, measured (**a**), at 370 K, (**b**), at 260 K, (**c**), at 180 K, and (**d**), at 130 K. The data were taken using the Disk-Chopper Time-of-Flight Spectrometer, DCS. Open white circles are the data, and the color coded surface is the surface image of the model calculated quasi-elastic intensity described in the text.

In order to visualize better the quasi-elastic and elastic signals, we integrated the data over three different energy ranges, $-0.1 < \hbar\omega < 0.1$ meV, $0.2 < \hbar\omega < 0.4$ meV, and $0.8 < \hbar\omega < 1.0$ meV and plotted their $Q$-dependences, $I(Q)$, in Fig. 2. For $-0.1 < \hbar\omega < 0.1$ meV (see Fig. 2 (a)), there are several sharp nuclear Bragg peaks. In addition, each $T$ data exhibit a prominent broad peak that is strong at low $Q$s and falls off as $Q$ increases. At high $Q$s ($Q \sim 2$ Å$^{-1}$), $I(Q)$ is almost zero for 370 K. However, the high $Q$ signal increases gradually upon cooling from the cubic phase (370 K) into the tetragonal phase (down to 180 K); see the data of 300 K (orange), 260 K (green) and 180 K (blue). Upon further cooling, $I(Q)$ becomes strong even at high $Q$s when the crystal structure becomes orthorhombic (130 K), resulting in its slow decrease with increasing $Q$; see the data in violet.

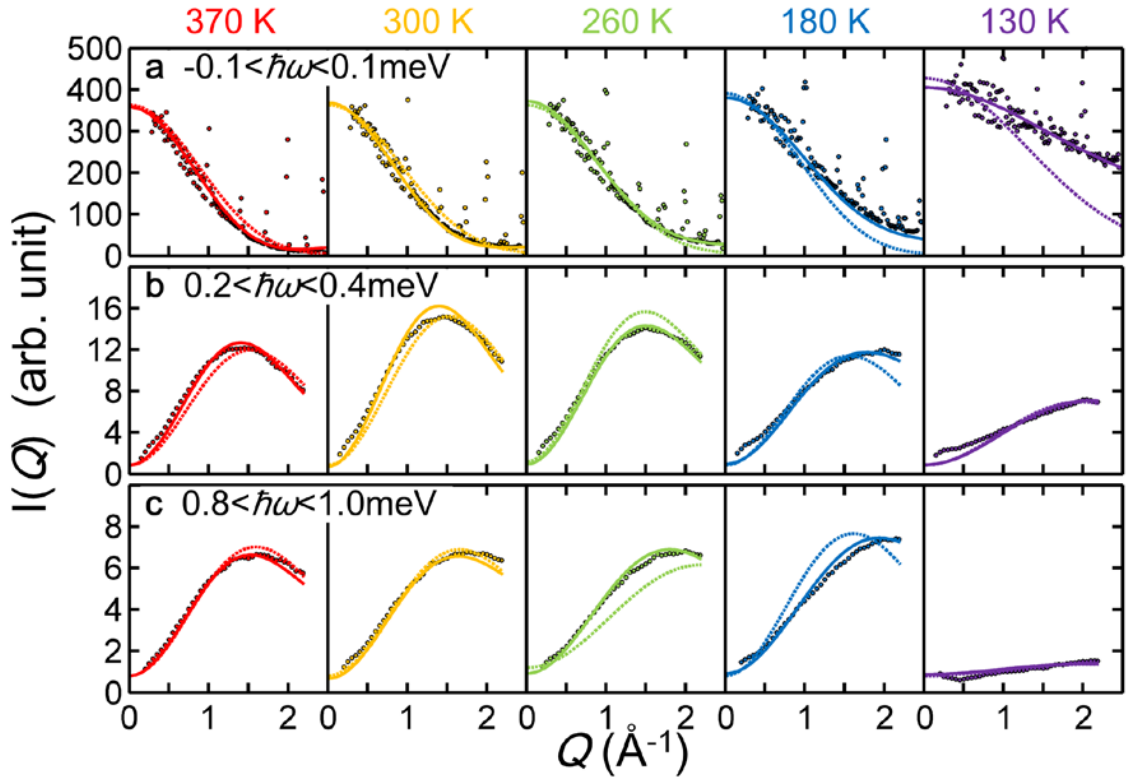

**Figure 2. Constant energy cuts of the CH$_3$NH$_3$PbI$_3$ neutron scattering data.** $Q$-dependences of the neutron scattering intensity for three different energy regions of (**a**), $-0.1 < \hbar\omega < 0.1$ meV, (**b**), $0.2 < \hbar\omega < 0.4$ meV, and (**c**), $0.8 < \hbar\omega < 1.0$ meV, are plotted for five different temperatures. Circles are the energy-integrated data, and the lines are the based on the $C_4 \otimes C_3$ jump model (solid) and isotropic rotational model (dotted)

as described in the text. The line is color-coded according to temperature; 370 K (red), 300 K (orange), 260 K (green), 180 K (blue) and 130 K (violet).

For the $0.2 < \hbar\omega < 0.4$ meV and $0.8 < \hbar\omega < 1$ meV energy windows (Fig. 2 (b) and (c), respectively), $I(Q)$ is peaked at $Q \sim 1.5$ Å$^{-1}$ at 370 K. Upon cooling down to 180 K, $I(Q)$ gradually shifts to higher $Q$ and becomes peaked at $Q \sim 2$ Å$^{-1}$ at 180 K. Upon further cooling, more prominent changes occur; at 130 K the peak position seems to have moved to $Q$ higher than 2 Å$^{-1}$, and $I(Q)$ becomes weak especially for $0.8 < \hbar\omega < 1$ meV. The weak intensity at 130 K over these energy windows is consistent with the strong enhancement of the intensity in the energy window of $-0.1 < \hbar\omega < 0.1$ meV (see Fig. 2 (a)). These indicate that dramatic changes occur when the system enters the orthorhombic phase.

**Models for Rotational Motions of CH$_3$NH$_3^+$**

Let us now explain the models that we have considered to explain the DCS data. As shown in Fig. S2 of the Electronic Supplementary Information, the quasi-elastic scattering is absent in CsPbI$_3$. This means that the scattering is due to motions of CH$_3$NH$_3^+$ that is located at the center of the cuboctahedral cage formed by the neighboring 12 iodide atoms. Due to the strong incoherent neutron scattering amplitude of hydrogen atoms, neutron scattering measurements provide a superb sensitivity toward probing dynamics of CH$_3$NH$_3^+$. By analyzing the $Q$ and $\hbar\omega$ dependent scattering intensities, the nature of the hydrogen motion can be determined.[47]

**An Isotropic Rotational Model for $CH_3NH_3^+$**

We have first considered an isotropic model for rotational motions of $CH_3NH_3^+$ [47]

$$S(Q,\hbar\omega) = e^{-\langle u^2\rangle Q^2}\left(j_0^2(Qr)\delta(\omega) + \sum_{l=1}^{\infty}(2l+1)j_l^2(Qr)\frac{1}{\pi}\frac{2\tau/l(l+1)}{1+\omega^2\left(\frac{2\tau}{l(l+1)}\right)^2}\right) \quad (1)$$

where $e^{-\langle u^2\rangle Q^2}$ is the Debye Waller factor, $\langle u^2\rangle$ is the mean squared displacement, $r$ is the radius of the object involved with the rotations, $j_l(Qr)$ is the $l$th spherical Bessel function, and $\tau_l = \frac{2\tau}{l(l+1)}$ is the relaxation time for the rotation with an angular momentum number $l$. This isotropic model describes hydrogen atoms randomly moving on the surface of a sphere. The best fits to the data are shown as the dotted lines in Fig. 2. For 370 K (cubic phase), the model fits reasonably well with $r_{iso} = 1.17(8)$ Å and $\tau = 1.62(2)$ ps. Terms with $l$ higher than 5 were ignored because $j_{l>5}^2(Qr)$ becomes negligible. $r_{iso}$ being close to the radius of the $CH_3NH_3^+$ molecule, 1.49(2) Å, suggests that the molecule undergoes nearly random rotations at 370 K. For 300 K in the tetragonal phase, the isotropic rotation model also reproduces the data reasonably well with $\tau = 2.33(3)$ ps. The increasing $\tau$ with decreasing $T$ is consistent with the expected energy softening of the rotational motions. For $T < 300$ K, however, the discrepancy between the model and the data becomes apparent; at 260 K and 180 K the model calculation falls off faster with increasing $Q$ than the data for $-0.1 < \hbar\omega < 0.1$ meV. Also, most notably, the model produces stronger intensity for $0.2 < \hbar\omega < 0.4$ meV and weaker scattering for $0.8 < \hbar\omega < 1$ meV than the data. The inadequacy of the isotropic model indicates that the rotational motions of $CH_3NH_3^+$ have some preferred directions in the tetragonal phase. At 130 K in the orthorhombic phase, the failure is clearly seen in $I(Q, -0.1 < \hbar\omega < 0.1$ meV).

**Introduction to Jump Models [47]**

The rotation model that accounts for the existence of a preferential molecular orientation is called a jump model. [47] Since the $CH_3NH_3^+$ molecule is located inside the cuboctahedral cage, its rotational motions are restricted by its own symmetry as well as by the local crystal symmetry of the cage. The possible rotational modes can be accounted for by the irreducible representations of the direct product of the symmetry of the local crystal environment ($C$) and that of the molecule ($M$); $\Gamma = C \otimes M$. Here we consider proper rotations only and do not consider improper rotations such as inversion and mirror reflection. Since this group theory has been extensively described in many textbooks including Ref. [47], we will state here only the basic formalism that is necessary for our discussion. In the group theory, the static and dynamic structure factor for rotational motions of molecules embedded in a crystal can be written as [47]

$$S(Q, \hbar\omega) = e^{-\langle u^2 \rangle Q^2} \left( \sum_\gamma A_\gamma(Q) \frac{1}{\pi} \frac{\tau_\gamma}{1 + \omega^2 \tau_\gamma^2} \right) \qquad (2)$$

where the sum over $\gamma$ runs over all the irreducible representations of $\Gamma$, $\Gamma_\gamma$. For a polycrystalline sample, $A_\gamma(Q)$ is given by

$$A_\gamma(Q) = \frac{l_\gamma}{g} \sum_\alpha \sum_\beta \chi_\gamma^{\alpha\beta} \sum_{C_\alpha} \sum_{M_\beta} j_0(Q|\boldsymbol{R} - C_\alpha M_\beta \boldsymbol{R}|) \qquad (3)$$

Here $g$ is the order of $\Gamma$ and $l_\gamma$ is the dimensionality of $\Gamma_\gamma$. The sums over $\alpha$ and $\beta$

run over all the classes of **C** and **M**, respectively, and the sums over $C_\alpha$ and $M_\beta$ run over all the rotations that belong to the crystal class, $\alpha$, and to the molecule class, $\beta$, respectively. The characters of $\boldsymbol{\Gamma}_\gamma$, $\chi_\gamma^{\alpha\beta}$, are the products of the characters of $\boldsymbol{C}_{\gamma_C}$ and $\boldsymbol{M}_{\gamma_M}$, $\chi_{\gamma_C}^\alpha$ and $\chi_{\gamma_M}^\beta$, respectively; $\chi_\gamma^{\alpha\beta} = \chi_{\gamma_C}^\alpha \chi_{\gamma_M}^\beta$. $j_0(x)$ is the zeroth spherical Bessel function and, $|\boldsymbol{R} - C_\alpha M_\beta \boldsymbol{R}|$, called the jump distance, is the distance between the initial atom position $\boldsymbol{R}$, and final atom position $C_\alpha M_\beta \boldsymbol{R}$. The relaxation time for the $\boldsymbol{\Gamma}_\gamma$ mode, $\tau_\gamma$, can be written in terms of the relaxation times for $\boldsymbol{C}_\alpha$ and $\boldsymbol{M}_\beta$, $\tau_\alpha$ and $\tau_\beta$, respectively,

$$\frac{1}{\tau_\gamma} = \sum_\alpha \frac{n_\alpha}{\tau_\alpha}\left(1 - \frac{\chi_\gamma^{\alpha e}}{\chi_\gamma^{Ee}}\right) + \sum_\beta \frac{n_\beta}{\tau_\beta}\left(1 - \frac{\chi_\gamma^{E\beta}}{\chi_\gamma^{Ee}}\right) \quad (4)$$

where $n_\alpha$ and $n_\beta$ are the number of symmetry rotations that belong to the classes, $\alpha$ and $\beta$, respectively. $E$ and $e$ represent the identity operations of **C** and **M**, respectively.

**Jump Model with $\boldsymbol{\Gamma} = \boldsymbol{C_4} \otimes \boldsymbol{C_3}$ for the Cubic and Tetragonal phases of $CH_3NH_3^+$**

Here we provide the simplest jump models that reproduce our DCS data. For both cubic and tetragonal phases, we used $\boldsymbol{\Gamma} = \boldsymbol{C} \otimes \boldsymbol{M} = \boldsymbol{C_4} \otimes \boldsymbol{C_3}$ where $\boldsymbol{C_4}$ and $\boldsymbol{C_3}$ represent the four-fold symmetry of the cuboctahedral cage and the three-fold symmetry of the molecule, respectively, as shown in Fig. 3 (b). According to a recent neutron diffraction study on the crystal structure,[48] the $CH_3NH_3^+$ cation axis lies in the *ab* plane in the tetragonal phase, with the long unique axis chosen to be *c*-axis $(c > a = b)$, leading to 4-fold and 2-fold symmetry about the *c*-axis.[48] The center of the $CH_3NH_3^+$ cation is slightly above the center of the cuboctahedral cage as determined by the neutron diffraction, and there is no two-fold symmetry about the *a*- and *b*-axis. Thus, the point group of the local crystal environment for the $CH_3NH_3^+$ cation, $\boldsymbol{C}$, is $\boldsymbol{C_4}$ rather than the full tetragonal group $D_4$. In the cubic phase, the center of the $CH_3NH_3^+$ cation moves to the center of the cage, and $\boldsymbol{C}$ becomes $O$ with preferred orientations along all the three principal *a*-, *b*-, and *c*-axes. Our study shows however that the structure factor of the $\boldsymbol{O \otimes C_3}$ model is basically identical to that of the $\boldsymbol{C_4 \otimes C_3}$ model for the $Q \lesssim 2\,\text{Å}^{-1}$ range covered by DCS. Neutron scattering measurements over much larger $Q$ range will be necessary in order to distinguish the two models. Thus we will use $C_4$ even for the cubic phase for simplicity and we will see that the model works well for the temperature range that is covered, $T \leq 370$ K.

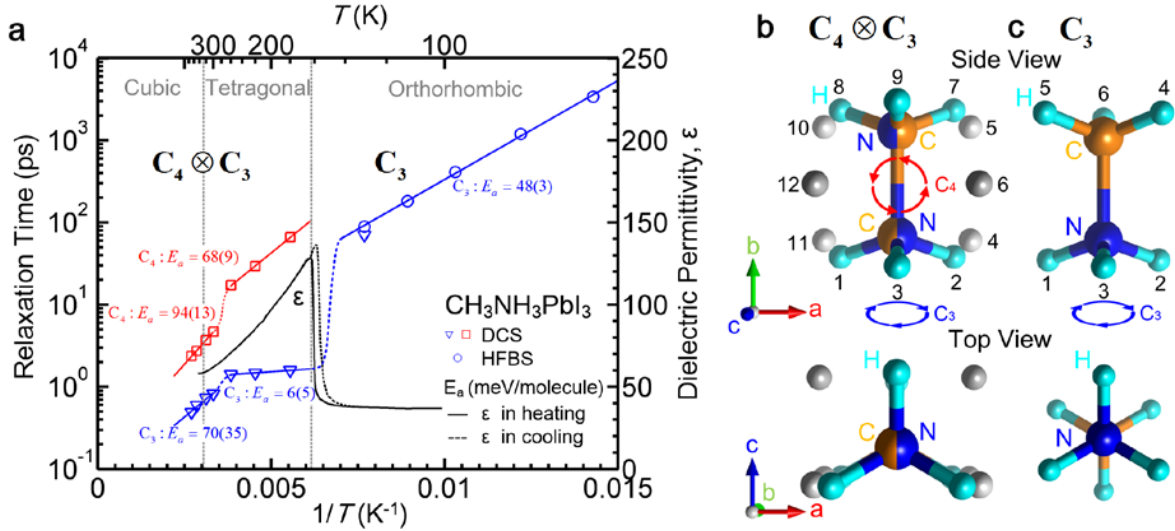

**Figure 3. Arrhenius plot for $CH_3NH_3PbI_3$ and the $CH_3NH_3^+$ cation.** (**a**) Relaxation time, $\tau$, of rotational modes of $CH_3NH_3PbI_3$ was obtained from fitting the model described in the text to the neutron scattering data at various temperatures that were taken at two different spectrometers, DCS (squares and triangles) and HFBS (circles). Some of these are shown in Fig. 2 and 3, and are plotted against inverse temperature (1/*T*). The colored solid lines are the fits to $\ln(\tau) = \frac{E_a}{k_B T} - \ln(A)$, where $E_a$ is an activation energy in unit of meV

per one molecule, $k_B$ the Boltzmann constant, and $A$ a pre-exponential factor. The colored dotted lines are guides to eyes. The black solid and dotted lines are temperature dependent dielectric permittivity measured by Onoda-Yamamuro *et al.*[43] which show a sharp increase at 160K (see the main text for more discussions) (**b**) The $CH_3NH_3^+$ molecule in the tetragonal environment. There are 12 crystallographically equivalent sites for hydrogen atoms. Since the motions of the two halves of $CH_3NH_3^+$ can be treated independently in the jump model calculations[49], the H atoms attached to C and N are plotted in an eclipsed configuration rather than the staggered configuration shown in (**c**), to clearly show the jump distances. (**c**) The $CH_3NH_3^+$ molecule in the orthorhombic environment. There are 3 crystallographically equivalent sites for each hydrogen atoms.

As described in detail in the Electronic Supplementary Information, the point group $\boldsymbol{C_4}$ has three irreducible representations, two one-dimensional (A and B) and one two-dimensional (E), and the point group $\boldsymbol{C_3}$ has two irreducible representations, one one-dimensional (A) and one two-dimensional (E). Thus, $\boldsymbol{\Gamma}$ has six irreducible representations; $\boldsymbol{\Gamma_\gamma} = \{A \otimes A, A \otimes E, B \otimes A, B \otimes E, E \otimes A, E \otimes E\}$. It is straightforward to calculate $A_\gamma(Q)$ and $\tau_\gamma$ using Eqs. (3) and (4) (see the Electronic Supplementary Information for details). As summarized in Table 1, there are three different relaxation times, $\tau_{C_4}, \tau_{C_2}, \tau_{C_3}$, that are associated with the symmetry operations, $C_4, C_2, C_3$, respectively. The $Q$-dependence of $A_\gamma(Q)$ was obtained from the jump distances $|\boldsymbol{R} - C_\alpha M_\beta \boldsymbol{R}|$ that were fixed to the values obtained from the crystal structure. The model calculated $S(Q, \hbar\omega)$ was convoluted with the instrumental energy resolution to fit the data. As shown as the color-coded surfaces in Fig. 1 and as the solid lines in Fig. 2, the $\boldsymbol{C_4} \otimes \boldsymbol{C_3}$ model reproduces the data extremely well over $180\text{ K} \leq T \leq 370\text{ K}$ spanning the cubic and tetragonal phases. We stress that the fit was carried out globally over the entire $Q$ and $\hbar\omega$ range with only four fitting parameters, an overall factor and the three relaxation times.

| $\Gamma_\gamma$ | $\dfrac{1}{\tau_\gamma}$ | $36 \cdot A_\gamma(Q)$ |
|---|---|---|
| A⊗A | 0 | $3 + 6j_1 + 4j_2 + 2j_3 + 2j_4 + 3j_5 + 2j_6 + 4j_7 + 4j_8 + 4j_9 + 2j_{10}$ |
| A⊗E | $\dfrac{3}{\tau_{C_3}}$ | $6 - 6j_1 + 8j_2 - 2j_3 + 4j_4 - 2j_6 - 4j_7 - 4j_8 - 4j_9 + 4j_{10}$ |
| B⊗A | $\dfrac{4}{\tau_{C_4}}$ | $3 + 6j_1 - 4j_2 - 2j_3 + 2j_4 + 3j_5 - 2j_6 - 4j_7 + 4j_8 - 4j_9 - 2j_{10}$ |
| B⊗E | $\dfrac{4}{\tau_{C_4}} + \dfrac{3}{\tau_{C_3}}$ | $6 - 6j_1 - 8j_2 + 2j_3 + 4j_4 + 2j_6 + 4j_7 - 4j_8 + 4j_9 - 4j_{10}$ |
| E⊗A | $\dfrac{2}{\tau_{C_4}} + \dfrac{2}{\tau_{C_2}}$ | $6 + 12j_1 - 4j_4 - 6j_5 - 8j_8$ |
| E⊗E | $\dfrac{2}{\tau_{C_4}} + \dfrac{2}{\tau_{C_2}} + \dfrac{3}{\tau_{C_3}}$ | $12 - 12j_1 - 8j_4 + 8j_8$ |

**Table 1. The jump model for $\Gamma = C_4 \otimes C_3$.** $j_i$ represents the zeroth spherical Bessel function $j_i = j_0(Qr_i)$. The jump distances $r_i$ are, using the notation $R_{i,j} = |R_i - R_j|$, $r_1 = R_{1,2}$, $r_2 = R_{1,4}$, $r_3 = R_{1,5}$, $r_4 = R_{1,7}$, $r_5 = R_{1,8}$, $r_6 = R_{1,11}$, $r_7 = R_{1,6}$, $r_8 = R_{1,9}$, $r_9 = R_{1,12}$, $r_{10} = R_{3,6}$ where $R_i$ is the position of the $i$th hydrogen atom numbered as in Fig. 3 (b).

The $C_4 \otimes C_3$ model is significantly better than the isotropic model at fitting the data at all temperatures, in particular at lower temperatures. This demonstrates that the $CH_3NH_3^+$ cation rotates with preferred orientations that are associated with the four-fold crystallographic $c$-axis, $C_4$, and the three-fold molecular symmetry around the C-N axis, $C_3$. Similar analysis was done for the DCS data taken at seven different temperatures, and the resulting relaxation times are listed in Table 3, and are plotted as a function of $1/T$ in a log scale in Fig. 3 (a) (square and triangular symbols). Upon cooling from 370 K (cubic phase), both $\ln(\tau_{C_4})$ and $\ln(\tau_{C_3})$ increase linearly with $1/T$ down to about 300 K. Upon further cooling well into the tetragonal phase, however, they exhibit a sudden jump to higher values after which they increase again linearly with $1/T$ until $\tau_{C_4}$ reaches 66(2) ps at 180 K. As the system transitions into the orthorhombic phase, the rotation about the four-fold $c$-axis freezes out.

**Jump Model with $\Gamma = C_3$ for the Orthorhombic Phase of $CH_3NH_3^+$**

In the orthorhombic phase, the orientation of the C-N axis is fixed due to hydrogen bonding between the $NH_3$ and iodide atoms and the resulting lattice distortion.[48] Thus $C = E$, and $\Gamma = E \otimes C_3 = C_3$. We have tried the $C_4 \otimes C_3$ model but the relaxation times associated with $C_4$ came out to be infinite indicating that the $C_4$ rotation is frozen as expected. Table 2 lists $A_\gamma(Q)$ and $\tau_\gamma$ that were calculated for the two irreducible representations of the $C_3$ point group using Eqs. (3) and (4) (see the Electronic Supplementary Information for details). As shown in Fig. 2, the $C_3$ model with $\tau_{C_3} = 70(6)$ ps reproduces the 130 K DCS data well.

| $\Gamma_\gamma$ | $\dfrac{1}{\tau_\gamma}$ | $3 \cdot A_\gamma(Q)$ |
|---|---|---|
| A | 0 | $1 + 2j_0(Qr)$ |
| E | $\dfrac{3}{\tau_{C_3}}$ | $2 - 2j_0(Qr)$ |

**Table 2. The jump model for $\Gamma = E \otimes C_3 = C_3$.** The 3-fold rotation jump distance is $r = |R_1 - R_2|$, where $R_i$ is the position of the $i$th hydrogen atom numbered as in Fig. 3 (c).

For $T < 130$ K, the high flux backscattering spectrometer, HFBS, with a superior energy resolution of 1 $\mu eV$ was used to study the slower rotation dynamics. Fig 4 shows the HFBS data taken at five different temperatures, 130 K (red), 112 K (orange), 97 K (green), 82 K (blue), and 70 K (violet). The data was fitted to a sum of a Lorentzian and a Gaussian to account for the broad quasi-elastic peak and the energy resolution limited elastic peak, respectively. The data shows that the quasi-elastic peak gets sharper as $T$ decreases, indicating increasing the relaxation time for the rotational motion. The relation between the Half-Width-Half-Maximum (HWHM) of the lorentzian and $\tau_{C_3}$ is given by HWHM ($\mu$eV) $= \frac{\hbar}{\tau_E} = \frac{3\hbar}{\tau_{C_3} \text{ (ps)}}$ where $\tau_E$ is the relaxation time for the motion associated with the two-dimensional $E$ representation of the $C_3$ point group (see Table 2). The best fit values for $\tau_{C_3}$ are listed for those temperatures in Table 3 and are plotted as circles in Fig. 3 (a).

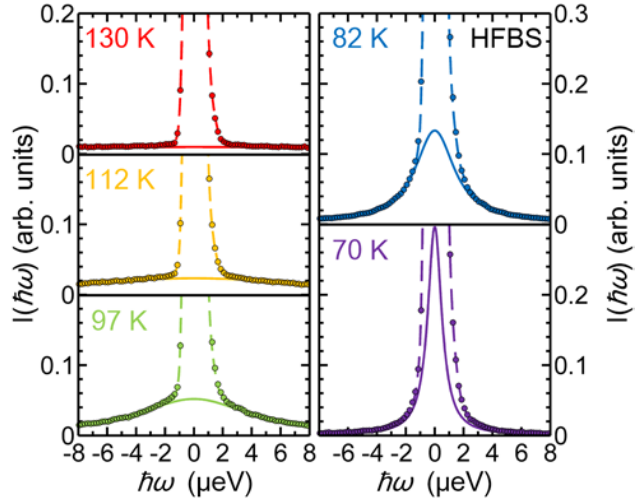

**Figure 4. Low energy scattering at $\mu$ eV level from CH$_3$NH$_3$PbI$_3$.** Neutron backscattering data were taken on HFBS at five different temperatures, 130K (red), 112 K (orange), 97 K (green), 82 K (blue), and 70 K (violet). The data were fitted to a sum of a lorentzian that accounts for the broad quasi-elastic peak and a gaussian for the energy resolution-limited elastic peak. The dashed and solid lines are the sum and the lorentzian, respectively.

| Phase | $T$ (K) | $\tau_{C_3}$ (ps) | $\tau_{C_4}$ (ps) | $\sqrt{\langle u^2 \rangle}$ (Å) | |
|---|---|---|---|---|---|
| Cubic | 370 | 0.49(2) | 2.37(2) | 0.27(5) | |
| | 350 | 0.60(2) | 2.72(2) | 0.28 (3) | |
| Tetragonal | 320 | 0.73(2) | 3.70 (4) | 0.30(2) | DCS |
| | 300 | 0.82(2) | 4.70 (6) | 0.31(2) | |
| | 260 | 1.42(2) | 17.3(3) | 0.35(2) | |
| | 220 | 1.48(7) | 30(2) | 0.34(3) | |
| | 180 | 1.59(1) | 66(2) | 0.33(7) | |
| Orthorhombic | 130 | 70(6) | N/A | 0.28(5) | |
| | 130 | 89(3) | N/A | N/A | HFBS |
| | 112 | 182(1) | N/A | N/A | |
| | 97 | 411(2) | N/A | N/A | |
| | 82 | 1195(6) | N/A | N/A | |
| | 70 | 3390(21) | N/A | N/A | |

**Table 3. Relaxation times $\tau_{C_4}$ and $\tau_{C_3}$ for the rotations of CH$_3$NH$_3$ about the *c*-axis and the C-N bond**, respectively. The values of $C_4$ and $C_3$ were obtained from fitting the DCS data to the $C_4 \otimes C_3$ model for 370 K $\leq T \leq$ 180 K and to the $C_3$ model for 130 K. For 370 K $\leq T \leq$ 180 K, $\tau_{C_2} \approx \infty$ for all temperatures. $\sqrt{\langle u^2 \rangle}$ is the square root of the mean squared displacement for the Debye Waller factor, $e^{-\langle u^2 \rangle Q^2}$. For $T < 130$ K, the quasi-elastic part of the HFBS data shown in Fig. 4 was fitted to a single lorentzian, and $\tau_{C_3}$ was obtained from the HWHM of the lorentzian assuming that HWHM ($\mu$eV) =

$\frac{3\hbar}{\tau_{C_3} \text{ (ps)}}$ where the factor 3 comes from the $C_3$ model (see Table 2). The errors in the parenthesis were estimated by the least square fitting with 95 % confidence.

## Discussion

Our group theoretical analysis of the neutron scattering data enabled us to separate the dynamics of the rotation of the C-N axis ($C_4$) from the rotation of the H atoms about the C-N axis ($C_3$). In the cubic phase ($T > 327$ K), our analysis yields $\tau_{C_4} \approx 3$ ps which is close to the previous experimental results obtained by millimeter-wave spectroscopy,[34] nuclear magnetic resonance,[40, 41] as well as theoretical prediction with *ab initio* molecular dynamics simulations.[18, 35, 42] Upon cooling, both $\tau_{C_4}$ and $\tau_{C_3}$ increase exponentially with $1/T$ (Figure 3a). Interestingly, both $\ln(\tau_{C_4})$ and $\ln(\tau_{C_3})$ undergo rapid increases whenever the system enters well into a new crystal structure with a lower symmetry. We can fit each linear region of $\ln(\tau)$ to the Arrhenius law, $\ln(\tau) = \frac{E_a}{k_B T} + \ln(\tau(T \to \infty))$, and obtain the activation energy $E_a$ for each rotational mode in each structural phase. The activation energy for the $C_4$ mode, $E_a(C_4)$, is 94(13) meV for the cubic phase and 68(9) meV for the tetragonal phase. The activation energy for the $C_3$ rotation, $E_a(C_3)$, also decreases from 70(35) meV in cubic phase to 6(5) meV in tetragonal phase. Interestingly, the jump in activation energy occurs somewhere between 260 – 300 K which is below the cubic-tetragonal transition temperature of ~330 K.[50] The cubic-to-tetragonal structural phase change in $CH_3NH_3PbI_3$ occurs through gradual tilting of the $PbI_6$ octahedra wherein the Pb-I2-Pb angle is the order parameter of the transition. Weller *et al*. has shown that

20there is initial rapid decrease in the Pb-I2-Pb angle from the initial 180 degree as $T$ is lowered from 330 K to ~ 280 K. As $T$ is lowered further, the distortion continues but with a much smaller rate as if the distortion has reached an equilibrium state of the tetragonal phase.[48] Our results show that the dynamics of the $CH_3NH_3^+$ ion immediately below the transition temperature is similar to that of in cubic phase. The local environment for the $CH_3NH_3^+$ including hydrogen bonding interaction with iodide atoms is still similar to the cubic structure since the distortion angle is still small. However, as the Pb-I2-Pb angle becomes ~165 degrees below 280 K, there is an abrupt change in the interaction between $CH_3NH_3^+$ and the lead iodide octahedral, and $E_a(C_4)$ exhibits a jump. A computational study has found that the orientation of the $CH_3NH_3^+$ ions and the lead iodide octahedral are strongly correlated[19] which supports our results in combination of those of Weller *et al.*[48]

As temperature is lowered below the tetragonal-orthorhombic transition temperature of 160 K, the $C_4$ rotation disappears while $C_3$ rotation still exists with $E_a(C_3)$ jumping up to 48(3) meV. The disappearance of $C_4$ mode is in agreement with the previous NMR study that showed that the C-N bond orientation gets frozen in the orthorhombic phase.[40] Also, this is in qualitative and quantitative agreement with computational results from Mattoni *et al.*[35] which show freezing of C-N bond rotation below 160K. It should be noted that the observed temperature dependence of $C_4$ mode (red symbols in Fig. 3a) coincides with that of dielectric permittivity, ε (black lines in Fig. 3a). Upon heating from low temperatures, there is a sudden increase in ε at 160K which is exactly the temperature at which the $C_4$ mode gets activated. Moreover, as the relaxation time of $C_4$ decreases upon further heating, ε also decreases. This can be understood as the effect of alignment of the C-N axis and the associated dipole to the external field through the $C_4$ rotation which increases the effective polarizability of $CH_3NH_3PbI_3$.

In contrast, a recent QENS work by Leguy et al.[39] does not detect freezing of $CH_3NH_3^+$ in orthorhombic phase without clear distinction among rotational modes. Moreover, their relaxation times of $14 \pm 3$ ps at room temperature and activation energies of 9.9 meV and 13.5 meV are different by several factors from our results as well as previous studies.[18, 34, 41] We think that this is due to the limited range of their data in energy of $-0.55\,\text{meV} < \hbar\omega < 0.55\,\text{meV}$ as well as Q-range of $0.18\,\text{Å}^{-1} \lesssim Q \lesssim 1.8\,\text{Å}^{-1}$ with only five data points which limited their analysis. In contrast to their work, our DCS data cover much larger $\hbar\omega$ and $Q$ ranges, $-3\,\text{meV} < \hbar\omega < 3\,\text{meV}$ and $0.11\,\text{Å}^{-1} \lesssim Q \lesssim 2.5\,\text{Å}^{-1}$ respectively, and our HFBS data have a superior energy resolution of 1 $\mu$eV, which is ideal for studying dynamics at low temperatures. We would like to emphasize that our comprehensive data and full group theoretical analysis allowed us to understand the nature of the two distinct $C_4$ and $C_3$ rotational modes; in the orthorhombic phase the $C_4$ mode of the C-N axis gets frozen but the $C_3$ rotation of the hydrogen atoms around the C-N axis still exists, which resolve the seemingly contradicting previous results by Leguy *et al*.[39] and other reports.[18, 34, 41]

Our results provide important insights on the microscopic mechanism of the high photovoltaic performance of $CH_3NH_3PbI_3$. For a semiconductor to exhibit high photovoltaic performance, it must have high exciton dissociation yield, low charge recombination rates and high mobilities. $CH_3NH_3PbI_3$ possesses a combination of these properties wherein $CH_3NH_3^+$ ions are thought to play critical roles. In terms of exciton binding energy, previous optical measurements have determined the upper bound to be in the range of ~50 meV assuming a dielectric constant of 6.5.[51-53] However, a recent computational study by Even *et al*.[36] has predicted that the appearance of the rotation of $CH_3NH_3^+$ ions and the associated jump in the dielectric constant above 160K can significantly lower the exciton binding energy to 1~10 meV. Indeed, several recent experimental studies have confirmed this prediction by performing more accurate measurement of the exciton binding energy in $CH_3NH_3PbI_3$ at room temperature to be in the range of few meV.[37, 38] Consistently, previous studies reported that the dielectric permittivity of $CH_3NH_3PbI_3$ abruptly jumps at 160K upon heating[43, 54] and strong excitonic features that are present below 160K become suddenly very weak above 160K.[51] Our results clearly show that the $C_4$ mode of the $CH_3NH_3^+$ plays a central role in these behaviors.

Another key reason for the high solar cell performance of $CH_3NH_3PbI_3$ is its carrier lifetime[22, 23, 55] that is longer than that of other high solar cell performance semiconductors such as GaAs, CdTe and CIGS.[56] Previous studies have shown that the charge recombination dynamics in $CH_3NH_3PbI_3$ under typical solar cell operating conditions are dominated by interaction of free electrons and holes[57, 58] which is consistent with the extremely low exciton binding energy.[36-38] An intriguing charge recombination behavior of $CH_3NH_3PbI_3$ is the extremely low bimolecular recombination rate - four orders of magnitude slower[55] than the limit set by Langevin theory.[59] This slow charge recombination is crucial in achieving the extremely long diffusion length observed in $CH_3NH_3PbI_3$[7-11] even with relatively modest charge mobilities[55, 56] and therefore can be considered to be one of the most important contributors to the high solar cell performance.

Microscopically, the slow charge recombination rate in $CH_3NH_3PbI_3$ can originate from the presence of the $CH_3NH_3^+$ ions with dipole that causes local fluctuations in energy landscape and nanoscale separation and localization of electrons and holes.[15, 17-19] Interestingly, our $E_a(C_4)$ value of 68(9) meV is remarkably close to the experimental charge recombination activation energy of 75(8) meV determined by time resolved microwave conductance and photoluminescence measurements in the temperature range of 165 K to 300 K.[23] The close match between the two activation energies suggests that the rotation of $CH_3NH_3^+$ ions may be the rate limiting step for the charge recombination process in $CH_3NH_3PbI_3$.

## Conclusions

We revealed in-depth and quantitative information on the motion of $CH_3NH_3^+$ as a function of temperature as well as crystal symmetry. The rotational rates and activations energies of the $CH_3NH_3^+$ determined in this work show close relations to various optoelectronic processes and properties reported in the literature such as temperature dependent charge recombination, exciton dissociation and dielectric constant. This suggests that the photovoltaic performance of hybrid perovskites can be tuned by changing the dynamics of organic cations by tuning the structure of the organic cation such as size and dipole moment. Understanding the motions of organic cations is therefore a prerequisite for rational and rapid progress toward photovoltaic device performance optimization as well as development of novel hybrid perovskites.

## Acknowledgements

J.J.C. acknowledges support from NASA VSGC New Investigator Award. B.I. is funded by 70NANB10H256 from NIST, U.S. Department of Commerce. This work utilized facilities supported in part by the National Science Foundation under Agreement No. DMR-0944772.

## Notes and References

1. W. S. Yang, J. H. Noh, N. J. Jeon, Y. C. Kim, S. Ryu, J. Seo and S. I. Seok, *Science*, 2015, **348**, 1234-1237.
2. H. Zhou, Q. Chen, G. Li, S. Luo, T.-b. Song, H.-S. Duan, Z. Hong, J. You, Y. Liu and Y. Yang, *Science*, 2014, **345**, 542-546.
3. A. Kojima, K. Teshima, Y. Shirai and T. Miyasaka, *J. Am. Chem. Soc.*, 2009, **131**, 6050–6051.
4. B. J. Foley, D. L. Marlowe, K. Sun, W. A. Saidi, L. Scudiero, M. C. Gupta and J. J. Choi, *Applied Physics Letters*, 2015, **106**, 243904.
5. Y. Yamada, T. Nakamura, M. Endo, A. Wakamiya and Y. Kanemitsu, *Appl. Phys. Express*, 2014, **7**, 032302.
6. H. J. Snaith, *J. Phys. Chem. Lett.*, 2013, **4**, 3623-3630.
7. A. Buin, P. Pietsch, J. Xu, O. Voznyy, A. H. Ip, R. Comin and E. H. Sargent, *Nano Lett.*, 2014, **14**, 6281-6286.
8. Q. Dong, Y. Fang, Y. Shao, P. Mulligan, J. Qiu, L. Cao and J. Huang, *Science*, 2015, **347**, 967-970.
9. D. Shi, V. Adinolfi, R. Comin, M. Yuan, E. Alarousu, A. Buin, Y. Chen, S. Hoogland, A. Rothenberger, K. Katsiev, Y. Losovyj, X. Zhang, P. A. Dowben, O. F. Mohammed, E. H. Sargent and O. M. Bakr, *Science*, 2015, **347**, 519-522.
10. S. D. Stranks, G. E. Eperon, G. Grancini, C. Menelaou, M. J. P. Alcocer, T. Leijtens, L. M. Herz, A. Petrozza and H. J. Snaith, *Science*, 2013, **342**, 341-344.
11. G. Xing, N. Mathews, S. Sun, S. S. Lim, Y. M. Lam, M. Grätzel, S. Mhaisalkar and T. C. Sum, *Science*, 2013, **342**, 344-347.
12. E. J. Juarez-Perez, R. S. Sanchez, L. Badia, G. Garcia-Belmonte, Y. S. Kang, I. Mora-Sero and J. Bisquert, *J. Phys. Chem. Lett.*, 2014, **5**, 2390-2394.
13. R. Gottesman, E. Haltzi, L. Gouda, S. Tirosh, Y. Bouhadana, A. Zaban, E. Mosconi and F. De Angelis, *J. Phys. Chem. Lett.*, 2014, **5**, 2662-2669.
14. Y. Kutes, L. Ye, Y. Zhou, S. Pang, B. D. Huey and N. P. Padture, *J. Phys. Chem. Lett.*, 2014, **5**, 3335-3339.
15. J. M. Frost, K. T. Butler, F. Brivio, C. H. Hendon, M. van Schilfgaarde and A. Walsh, *Nano Lett.*, 2014, **14**, 2584-2590.

16. S. Liu, F. Zheng, N. Z. Koocher, H. Takenaka, F. Wang and A. M. Rappe, *J. Phys. Chem. Lett.*, 2015, **6**, 693-699.
17. J. Ma and L.-W. Wang, *Nano Lett.*, 2015, **15**, 248-253.
18. E. Mosconi, C. Quarti, T. Ivanovska, G. Ruani and F. D. Angelis, *Phys. Chem. Chem. Phys.*, 2014, **16**, 16137-16144.
19. C. Quarti, E. Mosconi and F. De Angelis, *Chem. Mater.*, 2014, **26**, 6557-6569.
20. C. Quarti, E. Mosconi and F. D. Angelis, *Phys. Chem. Chem. Phys.*, 2015, **17**, 9394-9409.
21. R. S. Sanchez, V. Gonzalez-Pedro, J.-W. Lee, N.-G. Park, Y. S. Kang, I. Mora-Sero and J. Bisquert, *J. Phys. Chem. Lett.*, 2014, **5**, 2357-2363.
22. C. S. Ponseca, T. J. Savenije, M. Abdellah, K. Zheng, A. Yartsev, T. Pascher, T. Harlang, P. Chabera, T. Pullerits, A. Stepanov, J.-P. Wolf and V. Sundström, *J. Am. Chem. Soc.*, 2014, **136**, 5189-5192.
23. T. J. Savenije, C. S. Ponseca, L. Kunneman, M. Abdellah, K. Zheng, Y. Tian, Q. Zhu, S. E. Canton, I. G. Scheblykin, T. Pullerits, A. Yartsev and V. Sundström, *J. Phys. Chem. Lett.*, 2014, **5**, 2189-2194.
24. B. Chen, J. Shi, X. Zheng, Y. Zhou, K. Zhu and S. Priya, *J. Mater. Chem. A*, 2015, **3**, 7699-7705.
25. H.-W. Chen, N. Sakai, M. Ikegami and T. Miyasaka, *J. Phys. Chem. Lett.*, 2015, **6**, 164-169.
26. H.-S. Kim, S. K. Kim, B. J. Kim, K.-S. Shin, M. K. Gupta, H. S. Jung, S.-W. Kim and N.-G. Park, *J. Phys. Chem. Lett.*, 2015, **6**, 1729-1735.
27. J. Beilsten-Edmands, G. E. Eperon, R. D. Johnson, H. J. Snaith and P. G. Radaelli, *Applied Physics Letters*, 2015, **106**, 173502.
28. M. Coll, A. Gomez, E. Mas-Marza, O. Almora, G. Garcia-Belmonte, M. Campoy-Quiles and J. Bisquert, *J. Phys. Chem. Lett.*, 2015, **6**, 1408-1413.
29. J. M. Azpiroz, E. Mosconi, J. Bisquert and F. D. Angelis, *Energy Environ. Sci.*, 2015, **8**, 2118-2127.
30. W. Tress, N. Marinova, T. Moehl, S. M. Zakeeruddin, M. K. Nazeeruddin and M. Grätzel, *Energy Environ. Sci.*, 2015, **8**, 995-1004.
31. E. L. Unger, E. T. Hoke, C. D. Bailie, W. H. Nguyen, A. R. Bowring, T. Heumüller, M. G. Christoforo and M. D. McGehee, *Energy Environ. Sci.*, 2014, **7**, 3690-3698.
32. Z. Xiao, Y. Yuan, Y. Shao, Q. Wang, Q. Dong, C. Bi, P. Sharma, A. Gruverman and J. Huang, *Nat Mater*, 2015, **14**, 193-198.
33. H. J. Snaith, A. Abate, J. M. Ball, G. E. Eperon, T. Leijtens, N. K. Noel, S. D. Stranks, J. T.-W. Wang, K. Wojciechowski and W. Zhang, *J. Phys. Chem. Lett.*, 2014, **5**, 1511-1515.
34. A. Poglitsch and D. Weber, *The Journal of Chemical Physics*, 1987, **87**, 6373-6378.
35. A. Mattoni, A. Filippetti, M. I. Saba and P. Delugas, *J. Phys. Chem. C*, 2015, 17421-17428.
36. J. Even, L. Pedesseau and C. Katan, *J. Phys. Chem. C*, 2014, **118**, 11566-11572.
37. Q. Lin, A. Armin, R. C. R. Nagiri, P. L. Burn and P. Meredith, *Nat Photon*, 2015, **9**, 106-112.
38. A. Miyata, A. Mitioglu, P. Plochocka, O. Portugall, J. T.-W. Wang, S. D. Stranks, H. J. Snaith and R. J. Nicholas, *Nat Phys*, 2015, **11**, 582-587.

39. A. M. A. Leguy, J. M. Frost, A. P. McMahon, V. G. Sakai, W. Kochelmann, C. Law, X. Li, F. Foglia, A. Walsh, B. C. O'Regan, J. Nelson, J. T. Cabral and P. R. F. Barnes, *Nat Commun*, 2015, **6**, 7124.
40. O. Knop, R. E. Wasylishen, M. A. White, T. S. Cameron and M. J. M. V. Oort, *Can. J. Chem.*, 1990, **68**, 412-422.
41. R. E. Wasylishen, O. Knop and J. B. Macdonald, *Solid State Communications*, 1985, **56**, 581-582.
42. M. A. Carignano, A. Kachmar and J. Hutter, *J. Phys. Chem. C*, 2015, **119**, 8991-8997.
43. N. Onoda-Yamamuro, T. Matsuo and H. Suga, *Journal of Physics and Chemistry of Solids*, 1992, **53**, 935-939.
44. M. M. Lee, J. Teuscher, T. Miyasaka, T. N. Murakami and H. J. Snaith, *Science*, 2012, **338**, 643-647.
45. J. R. D. Copley and J. C. Cook, *Chemical Physics*, 2003, **292**, 477-485.
46. A. Meyer, R. M. Dimeo, P. M. Gehring and D. A. Neumann, *Review of Scientific Instruments*, 2003, **74**, 2759-2777.
47. M. Bée, *Quasielastic Neutron Scattering*, Adam Hilger, Bristol, 1988.
48. M. T. Weller, O. J. Weber, P. F. Henry, A. M. D. Pumpo and T. C. Hansen, *Chem. Commun.*, 2015, **51**, 4180-4183.
49. J. Tsau and D. F. R. Gilson, *Can. J. Chem.*, 1970, **48**, 717-722.
50. T. Baikie, Y. Fang, J. M. Kadro, M. Schreyer, F. Wei, S. G. Mhaisalkar, M. Graetzel and T. J. White, *J. Mater. Chem. A*, 2013, **1**, 5628-5641.
51. V. D'Innocenzo, G. Grancini, M. J. P. Alcocer, A. R. S. Kandada, S. D. Stranks, M. M. Lee, G. Lanzani, H. J. Snaith and A. Petrozza, *Nat Commun*, 2014, **5**, 3586.
52. M. Hirasawa, T. Ishihara, T. Goto, K. Uchida and N. Miura, *Physica B: Condensed Matter*, 1994, **201**, 427-430.
53. K. Tanaka, T. Takahashi, T. Ban, T. Kondo, K. Uchida and N. Miura, *Solid State Communications*, 2003, **127**, 619-623.
54. O. Almora, I. Zarazua, E. Mas-Marza, I. Mora-Sero, J. Bisquert and G. Garcia-Belmonte, *J. Phys. Chem. Lett.*, 2015, **6**, 1645-1652.
55. C. Wehrenfennig, G. E. Eperon, M. B. Johnston, H. J. Snaith and L. M. Herz, *Adv. Mater.*, 2014, **26**, 1584-1589.
56. S. D. Stranks and H. J. Snaith, *Nat Nano*, 2015, **10**, 391-402.
57. K. Chen, A. J. Barker, F. L. C. Morgan, J. E. Halpert and J. M. Hodgkiss, *J. Phys. Chem. Lett.*, 2015, **6**, 153-158.
58. J. S. Manser and P. V. Kamat, *Nat Photon*, 2014, **8**, 737-743.
59. P. Langevin, *Annales de Chimie et de Physique*, 1903, **28**, 433-530.

# Electronic Supplementary Information for

# Rotational Dynamics of Organic Cations in CH$_3$NH$_3$PbI$_3$ Perovskite

Tianran Chen, Benjamin J. Foley, Bahar Ipek, Madhusudan Tyagi, John R. D. Copley, Craig M. Brown, Joshua J. Choi*, Seung-Hun Lee*

*Corresponding authors. E-mail: shlee@virginia.edu, jjc6z@virginia.edu

**Neutron Diffraction Data and Crystal Structure Refinement**

Fig. S1 (a) shows the neutron diffraction data taken at 4 K. The Bragg peaks were fitted using the Rietveld refinement method to determine the crystal structure at 5 K, using GSAS (*S1*) package along with EXPGUI (graphical user interface) (*S2*). The refined crystal structure with an orthorhombic *Pnma* space group was the same as that reported by Baikie *et al*. (*S3*) and Stoumpos *et al*. (*S4*) The structural parameters are listed in Table S1 and the resulting crystal structure is plotted in Fig. S1.

**Table S1.** Crystal structural parameters of CH$_3$NH$_3$PbI$_3$ for 4 K. Positions within orthorhombic space group ***Pnma*** and occupancies per a chemical unit cell (c.u.) in CH$_3$NH$_3$PbI$_3$ at $T = 4$ K as determined by Rietveld analysis of the neutron diffraction data shown in Fig. S1 using the GSAS. The lattice parameters are $a = 8.8155(4)$ Å, $b = 12.5980(6)$ Å, and $c = 8.5637(5)$ Å. Isotropic Debye-Waller factors, $\exp(-\langle u^2 \rangle Q^2)$, were used where $\langle u^2 \rangle$ is the mean squared displacement. The resulting overall reduced $\chi^2 = 1.089$.

| Site | $x$ | $y$ | $z$ | $n$/c.u. | $\sqrt{\langle u^2 \rangle}$/Å |
|---|---|---|---|---|---|
| Pb | 1/2 | 0 | 0 | 4 | 0.0095(9) |
| I1 | 0.4844(11) | 1/4 | -0.0531(10) | 4 | 0.0094(17) |
| I2 | 0.1871(6) | 0.0165(5) | 0.1818(6) | 8 | 0.0078(11) |
| N | 0.9390(6) | 3/4 | 0.0294(6) | 4 | 0.0127(11) |
| C | 0.9166(9) | 1/4 | 0.0683(10) | 4 | 0.0200(20) |
| H1 | 0.3536(14) | 0.1785(8) | 0.4603(17) | 8 | 0.0386(32) |
| H2 | 0.6305(10) | 0.1877(7) | 0.4983(14) | 8 | 0.0172(22) |
| H3 | 0.5506(28) | 1/4 | 0.6524(14) | 4 | 0.0370(46) |
| H4 | 0.4601(37) | 1/4 | 0.3159(19) | 4 | 0.0635(62) |

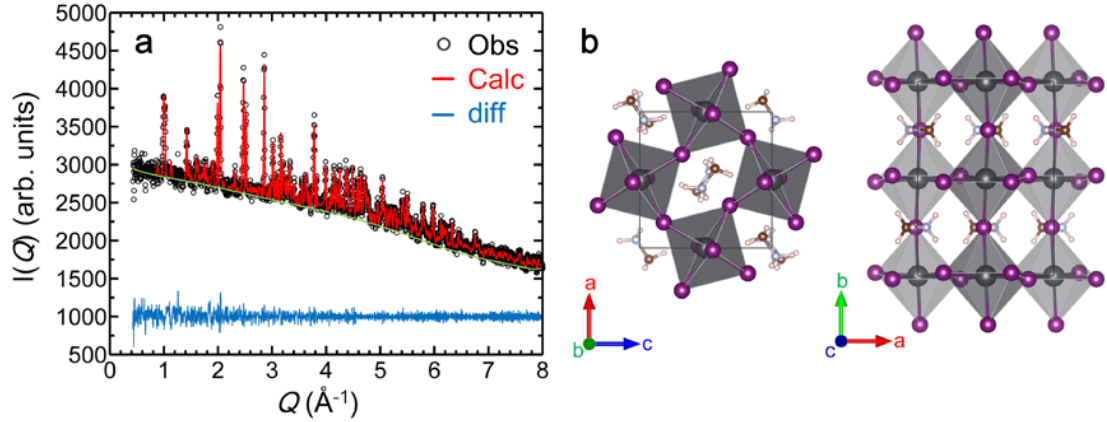

**Fig. S1** (**a**) Neutron diffraction data collected for $CH_3NH_3PbI_3$ sample at 4 K. (**b**) The crystal structure determined by Rietveld refinement of the NPD data. The structural parameters are listed in Table S1. Note that in the orthorhombic space group $Pnma$, the longest axis is chosen to be $b$-axis ($b = 12.5980(6)$ Å), while in the tetragonal space group $I4/mcm$, the long axis is chosen to be $c$-axis ($c = 12.5944(7)$ Å) in the literature (*S3-5*).

**DCS Data from $CH_3NH_3PbI_3$ and $CsPbI_3$**

Fig. S2 shows DCS data obtained from the powder samples of $CH_3NH_3PbI_3$ and $CsPbI_3$ at 300 K. As shown in Fig. S2 (a), $CH_3NH_3PbI_3$ exhibits strong quasi-elastic scattering at 300 K that is extended up to a few meV and is broad in $Q$. On the other hand, the strong quasi-elastic scattering is absent in $CsPbI_3$ at 300 K. Thus we conclude that the quasi-elastic scattering present in MAPbI$_3$ comes from motions of the $CH_3NH_3^+$, which is confirmed by our analysis described in the text.

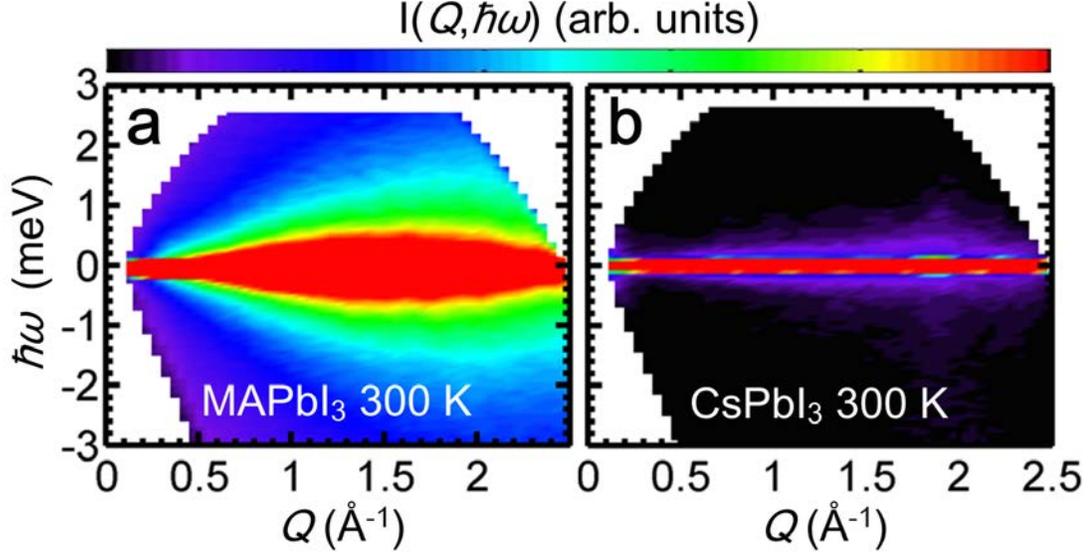

**Fig. S2.** Contour maps of quasi-elastic neutron scattering data taken from $CH_3NH_3PbI_3$ and $CsPbI_3$ at 300K at DCS $(\lambda = 4.8\ \text{Å})$.

**Jump Models for $CH_3NH_3PbI_3$**

(This section is in the main text but is repeated here for reader convenience.)

The rotation model that accounts for the existence of a preferential orientation is called a 'jump model'. (*S6*) Since the $CH_3NH_3^+$ molecule is located inside the cuboctahedral cage, its rotational motions are restricted by its own symmetry as well as by the local crystal symmetry of the cage. The possible rotational modes can be accounted for by the irreducible representations of the direct product of the symmetry of the local crystal environment ($C$) and that of the molecule ($M$); $\boldsymbol{\Gamma = C \otimes M}$. Here we consider proper rotations only and do not consider improper rotations such as inversion and mirror reflection. Since this group theory has been extensively described in many textbooks including Ref. (*S6*), we will state here only the basic formalism that is necessary for our discussion. In the group theory, the static and dynamic structure factor for rotational motions of molecules embedded in a crystal can be written as (*S6*)

$$S(Q, \hbar\omega) = e^{-\langle u^2 \rangle Q^2} \left( \sum_\gamma A_\gamma(Q) \frac{1}{\pi} \frac{\tau_\gamma}{1 + \omega^2 \tau_\gamma^2} \right) \qquad (S1)$$

where the sum over $\gamma$ runs over all the irreducible representations of $\Gamma$, $\Gamma_\gamma$. For a polycrystalline sample, $A_\gamma(Q)$ is given by (S6)

$$A_\gamma(Q) = \frac{l_\gamma}{g} \sum_\alpha \sum_\beta \chi_\gamma^{\alpha\beta} \sum_{C_\alpha} \sum_{M_\beta} j_0(Q|\mathbf{R} - C_\alpha M_\beta \mathbf{R}|) \qquad (S2)$$

Here $g$ is the order of $\Gamma$ and $l_\gamma$ is the dimensionality of $\Gamma_\gamma$. The sums over $\alpha$ and $\beta$ run over all the classes of $\mathbf{C}$ and $\mathbf{M}$, respectively, and the sums over $C_\alpha$ and $M_\beta$ run over all the rotations that belong to the crystal class, $\alpha$, and to the molecule class, $\beta$, respectively. The characters of $\Gamma_\gamma$, $\chi_\gamma^{\alpha\beta}$, are the products of the products of the characters of $\mathbf{C}_{\gamma_C}$ and $\mathbf{M}_{\gamma_M}$, $\chi_{\gamma_C}^\alpha$ and $\chi_{\gamma_M}^\beta$, respectively; $\chi_\gamma^{\alpha\beta} = \chi_{\gamma_C}^\alpha \chi_{\gamma_M}^\beta$. $j_0(x)$ is the zeroth spherical Bessel function and, $|\mathbf{R} - C_\alpha M_\beta \mathbf{R}|$, called the jump distance, is the distance between the initial atom position $\mathbf{R}$, and final atom position $C_\alpha M_\beta \mathbf{R}$. The relaxation time for the $\Gamma_\gamma$ mode, $\tau_\gamma$, can be written in terms of the relaxation times for $C_\alpha$ and $M_\beta$, $\tau_\alpha$ and $\tau_\beta$, respectively, (S6)

$$\frac{1}{\tau_\gamma} = \sum_\alpha \frac{n_\alpha}{\tau_\alpha}\left(1 - \frac{\chi_\gamma^{\alpha e}}{\chi_\gamma^{Ee}}\right) + \sum_\beta \frac{n_\beta}{\tau_\beta}\left(1 - \frac{\chi_\gamma^{E\beta}}{\chi_\gamma^{Ee}}\right) \qquad (S3)$$

where $n_\alpha$ and $n_\beta$ are the number of symmetry rotations that belong to the classes, $\alpha$ and $\beta$, respectively. $E$ and $e$ represent the identity operations of $\mathbf{C}$ and $\mathbf{M}$, respectively.

.

### $\Gamma = C_4 \otimes C_3$ Jump Model

In this model, the local crystal environment has a four-fold symmetry and the molecule itself has a three-fold rotational symmetry as the $CH_3NH_3^+$ does in the tetragonal phase (see Fig. S3). The character tables for the irreducible representations of the $C_4$ and $C_3$ point groups are shown in Table S2.

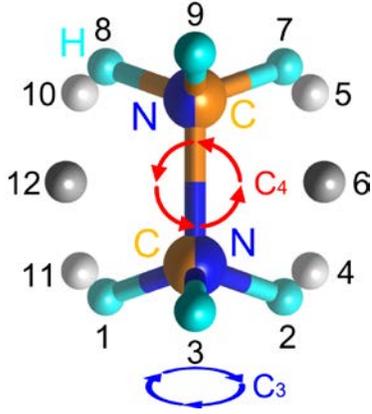

Fig. S3. The $CH_3NH_3^+$ in the tetragonal environment. The small spheres in sky blue and their counter parts in grey, representing 12 crystallographically equivalent sites for hydrogen atoms in the initial orientation and upon rotation of the molecular axis by 90°, respectively. Since the motions of the two halves of the $CH_3NH_3^+$ ion can be treated independently in the jump model calculations, the H atoms attached to C and N are plotted in an eclipsed configuration rather than the staggered configuration shown in Fig. S4, to clearly show the jump distances.

Table S2. The $C_4$ point group has three classes and three irreducible representations, two one-dimensional (A and B) and one two-dimensional (E) representations. The $C_3$ point group has two classes and two irreducible representations, a one-dimensional (A) and one two-dimensional (E) representations.

| $C = C_4$ | $E$ | $2C_4$ | $C_2$ |
|---|---|---|---|
| A | 1 | 1 | 1 |
| B | 1 | -1 | 1 |
| E | 2 | 0 | -2 |

| $M = C_3$ | $E$ | $2C_3$ |
|---|---|---|
| A | 1 | 1 |
| E | 2 | -1 |

The resulting point group for the possible rotational motions, $\Gamma$, has 6 irreducible representations: $\Gamma_\gamma = \{A \otimes A, A \otimes E, B \otimes A, B \otimes E, E \otimes A, E \otimes E\}$. It is straightforward to get correlation times from Eq. S3;

$$\frac{1}{\tau_{A \otimes A}} = 0, \frac{1}{\tau_{A \otimes E}} = \frac{3}{\tau_{C_3}}, \frac{1}{\tau_{B \otimes A}} = \frac{4}{\tau_{C_4}},$$

$$\frac{1}{\tau_{B \otimes E}} = \frac{4}{\tau_{C_4}} + \frac{3}{\tau_{C_3}}, \frac{1}{\tau_{E \otimes A}} = \frac{4}{\tau_{C_4}} + \frac{2}{\tau_{C_2}}, \frac{1}{\tau_{E \otimes E}} = \frac{2}{\tau_{C_4}} + \frac{2}{\tau_{C_2}} + \frac{3}{\tau_{C_3}},$$

Now let us calculate $A_\gamma(Q)$ for $\boldsymbol{\Gamma} = \boldsymbol{C_4} \otimes \boldsymbol{C_3}$. First, we need to know how a hydrogen atom moves from one to another crystallographically equivalent site under the combined symmetry operations of the two point groups of $\boldsymbol{C} = \boldsymbol{C_4}$ and $\boldsymbol{M} = \boldsymbol{C_3}$. Table S4 lists the geometrical information.

**Table S4.** Geometrical information for a hydrogen atom moving from one site to another equivalent site under the combined symmetry operations of the two point groups of $\boldsymbol{C} = \boldsymbol{C_4}$ and $\boldsymbol{M} = \boldsymbol{C_3}$. The resulting jump distances $r_i$ are, using the notation $R_{i,j} = |\boldsymbol{R_i} - \boldsymbol{R_j}|$, $r_1 = R_{1,2}$, $r_2 = R_{1,4}$, $r_3 = R_{1,5}$, $r_4 = R_{1,7}$, $r_5 = R_{1,8}$, $r_6 = R_{1,11}$, $r_7 = R_{1,6}$, $r_8 = R_{1,9}$, $r_9 = R_{1,12}$, $r_{10} = R_{3,6}$ where $\boldsymbol{R_i}$ is the position of the $i$th hydrogen atom numbered as in Fig. S3.

| Initial positions | | Final positions | | | Jump Distance | | | |
|---|---|---|---|---|---|---|---|---|
| | | $E$ | $C_3$ | $C_3^2$ | $E$ | $C_3$ | $C_3^2$ | |
| 1 | $E$ | 1 | 2 | 3 | 0 | $r_1$ | $r_1$ | $r_1 = R_{1,2}$ |
| | $C_4$ | 4 | 5 | 6 | $r_2$ | $r_3$ | $r_7$ | $r_2 = R_{1,4}$ |
| | $C_2$ | 7 | 8 | 9 | $r_4$ | $r_5$ | $r_8$ | $r_3 = R_{1,5}$ |
| | $C_4^3$ | 10 | 11 | 12 | $r_2$ | $r_6$ | $r_9$ | $r_4 = R_{1,7}$ |
| 2 | $E$ | 2 | 3 | 1 | 0 | $r_1$ | $r_1$ | $r_5 = R_{1,8}$ |
| | $C_4$ | 5 | 6 | 4 | $r_2$ | $r_3$ | $r_7$ | $r_6 = R_{1,11}$ |
| | $C_2$ | 8 | 9 | 7 | $r_4$ | $r_5$ | $r_8$ | $r_7 = R_{1,6}$ |
| | $C_4^3$ | 11 | 12 | 10 | $r_2$ | $r_6$ | $r_9$ | $r_8 = R_{1,9}$ |
| 3 | $E$ | 3 | 1 | 2 | 0 | $r_1$ | $r_1$ | $r_9 = R_{1,12}$ |
| | $C_4$ | 6 | 4 | 5 | $r_{10}$ | $r_9$ | $r_7$ | $r_{10} = R_{3,6}$ |
| | $C_2$ | 9 | 7 | 8 | $r_5$ | $r_8$ | $r_8$ | |
| | $C_4^3$ | 12 | 10 | 11 | $r_{10}$ | $r_9$ | $r_7$ | |

We can now calculate $A_\gamma(Q)$ using Eq. S3:

$$A_{A\otimes A}(Q) = \frac{1}{36}[3 + 6j_1 + 4j_2 + 2j_3 + 2j_4 + 3j_5 + 2j_6 + 4j_7 + 4j_8 + 4j_9 + 2j_{10}]$$

$$A_{A\otimes E}(Q) = \frac{1}{36}[6 - 6j_1 + 8j_2 - 2j_3 + 4j_4 - 2j_6 - 4j_7 - 4j_8 - 4j_9 + 4j_{10}]$$

$$A_{B\otimes A}(Q) = \frac{1}{36}[3 + 6j_1 - 4j_2 - 2j_3 + 2j_4 + 3j_5 - 2j_6 - 4j_7 + 4j_8 - 4j_9 - 2j_{10}]$$

$$A_{B\otimes E}(Q) = \frac{1}{36}[6 - 6j_1 - 8j_2 + 2j_3 + 4j_4 + 2j_6 + 4j_7 - 4j_8 + 4j_9 - 4j_{10}]$$

$$A_{E\otimes A}(Q) = \frac{1}{36}[6 + 12j_1 - 4j_4 - 6j_5 - 8j_8]$$

$$A_{E\otimes E}(Q) = \frac{1}{36}[12 - 12j_1 - 8j_4 + 8j_8]$$

where $j_i = j_0(Qr_i)$ is the zeroth spherical Bessel function.

The elastic and quasielastic incoherent structure factors, $A_\gamma(Q)$, are plotted in Fig. S4. Atomic positions of $CH_3NH_3^+$ are obtained from our refinement of 4 K neutron diffraction data.

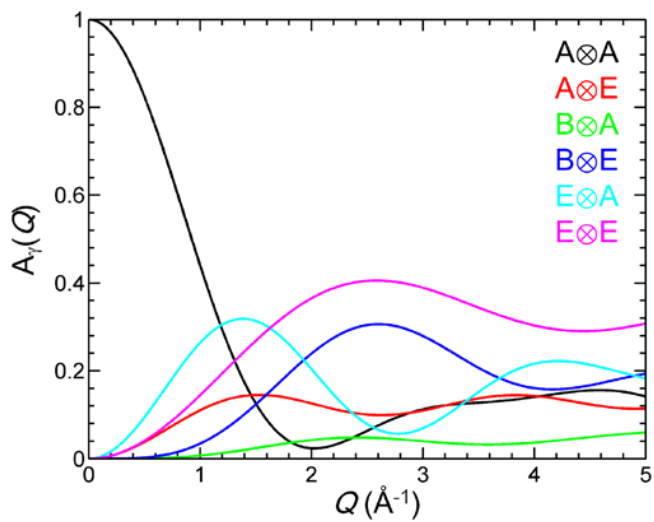

**Fig. S4.** $Q$-dependences of the calculated elastic and quasi-elastic incoherent structure factors for the $C_4 \otimes C_3$ jump model. The irreducible representation A⊗A gives elastic contribution $\left(\frac{1}{\tau_{A\otimes A}} = 0\right)$ while the rest give quasi-elastic contributions.

$\Gamma = E \otimes C_3 = C_3$ **Jump Model**

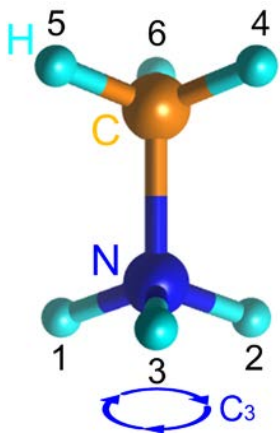

**Fig. S5.** $CH_3NH_3^+$ in the orthorhombic environment.

In this model, the orientation of the $CH_3NH_3^+$ is fixed due to hydrogen bonding between the $NH_3$ and iodide atoms. Thus $C = E$, and $\Gamma = E \otimes C_3 = C_3$. The character table of the point group $C_3$ is:

| $\Gamma = C_3$ | $E$ | $2C_3$ |
|---|---|---|
| A | 1 | 1 |
| E | 2 | -1 |

The correlation times can be obtained by,

$$\frac{1}{\tau_\gamma} = \sum_\alpha \frac{n_\alpha}{\tau_\alpha}\left(1 - \frac{\chi_\gamma^\alpha}{\chi_\gamma^E}\right) \qquad (S4)$$

It is straightforward to calculate the equation to get

$$\frac{1}{\tau_A} = 0, \frac{1}{\tau_E} = \frac{3}{\tau_{C_3}}$$

The elastic and quasi-elastic incoherent structure factors can be obtained by

$$A_\gamma(Q) = \frac{\chi_\gamma^E}{g} \sum_\alpha \chi_\gamma^\alpha \sum_{\Gamma_\alpha} j_0(Q|\mathbf{R} - \Gamma_\alpha \mathbf{R}|) \qquad (S5)$$

The jump distance of the three-fold rotation is $r = |\mathbf{R}_1 - \mathbf{R}_2|$, where $\mathbf{R}_i$ is the position of the $i$th H atom numbered as in Fig. S4. Thus, $A_\gamma(Q)$ becomes

$$A_A(Q) = \frac{1}{3}[1 + 2j_0(Qr)]$$

$$A_E(Q) = \frac{2}{3}[1 - j_0(Qr)]$$

The elastic and quasi-elastic incoherent structure factors, $A_\gamma(Q)$, are plotted as a function of $Q$ in Fig. S6.

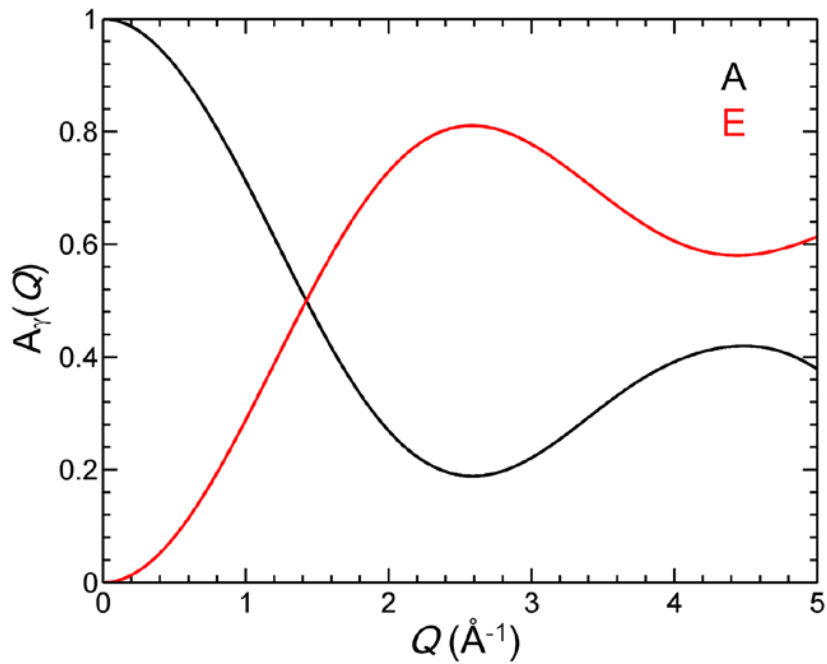

**Fig. S6.** $Q$-dependences of the calculated elastic and quasi-elastic incoherent structure factors for the $C_3$ jump model. The irreducible representation $A$ gives elastic contribution $\left(\frac{1}{\tau_{A\otimes A}} = 0\right)$ while $E$ gives quasi-elastic contribution.

## References

(*S1*) A. C. Larson and R. B. Von Dreele, Los Alamos National Laboratory, 1994; pp. 86–748.
(*S2*) B. H. Toby, *J. Appl. Crystallogr.* 2001, 34, 210–213.
(*S3*) T. Baikie, Y. Fang, J. M. Kadro, M. Schreyer, F. Wei, S. G. Mhaisalkar, M. Gratzel and T. J. White, *J. Mater. Chem. A* 2013, 1, 5628.
(*S4*) C. C. Stoumpos, C. D. Malliakas and M. G. Kanatzidis, *Inorg. Chem.* 2013, 52, 9019–9038.
(*S5*) M. T. Weller, P. Henry, O. Weber, A. Di Pumpo and T. Hansen, *Chem. Commun.*, 2015, DOI: 10.1039/C4CC09944C.
(*S6*) M. Beé, *Quasielastic Neutron Scattering* (Adam Hilger, Bristol, 1988).
(*S7*) Certain commercial equipment, instruments, or materials are identified in this paper to foster understanding. Such identification does not imply recommendation or endorsement by the National Institute of Standards and Technology, nor does it imply that the materials or equipment identified are necessarily the best available for the purpose.